\documentstyle[12pt]{article}
\begin{document}
\begin{titlepage}
\begin{flushright} UTPT-93-27 \\ hep-ph/9311287 \\ November 1993
\end{flushright}
\vspace{24pt}
\begin{center} {\LARGE $1/m_c$ Expansion of $\Omega_b \rightarrow
\Omega_c^{(\ast)}$
\\ Semileptonic Decay Form Factors}\\ \vspace{40pt} {\large M.
Sutherland}
\vspace{0.5cm}

Department of Physics, University of Toronto\\ Toronto,
Ontario, Canada M5S 1A7  \\ tel.: (416) 978-0712; fax: (416) 978-2537  \\
email: marks@medb.physics.utoronto.ca

\vspace{12pt}

\begin{abstract} A simple model of $\Omega_Q$ and $\Omega_Q^{\ast}$
baryons containing one heavy quark $Q$ is constructed.  Amplitudes are
represented by loop graphs with one line for the heavy quark and
another for the light degrees of freedom.  The latter are
modelled as a freely-propagating vector particle interacting nonlocally
with the heavy baryon and a free heavy quark.  It is argued that the physics
of confinement plays an inessential role in determining semileptonic decay
form factors.  The model has a well-defined heavy-quark expansion which has
a form consistent (through order $1/m_Q$) with that determined by QCD through
the heavy-quark effective theory.  The slope of the Isgur-Wise function is
consistent with the Bjorken sum rule bound.  The effect of the
$\Omega_Q^{\ast}-\Omega_Q$ mass splitting on the first-order form factors is
examined.
\end{abstract}\end{center} \end{titlepage}
\pagebreak

There is presently a large amount of theoretical interest in properties of
baryons with one heavy quark $Q$.  This is because the spin-flavor
symmetries of QCD in the limit $m_Q\rightarrow \infty$ give rise to
relations between certain weak decay form factors and to model-independent
absolute normalizations for others at zero recoil \cite{IW,Geo}.  More
generally, the heavy-quark effective theory (HQET) may be used to derive
relations among the many {\it a priori} independent corrections to the
form factors at any order in the expansion in inverse powers of heavy
quark masses. (For a review, see \cite{Neurevue}.) This information
serves to organize and classify the unknown quantities of QCD, but to be of
real use it remains necessary to have a model with which to perform
computations.  Of course, the QCD relations must be checked in any
particular model for consistency.  One case in which such considerations
place constraints on model parameters is described in \cite{NeuRieck}.
The purpose of this paper is to define a model for $\Omega_Q$ and
$\Omega_Q^{(\ast)}$ baryons and to show that it is consistent with the
form of the heavy-quark expansion in QCD at order $1/m_Q$ as derived in
\cite{BB}, with no extra constraints on its parameters.

In the heavy-quark limit, the $\Omega_Q$ and $\Omega_Q^{(\ast)}$ are the
$s=1/2$ and $s=3/2$ combinations, respectively, of a heavy quark $Q$ and
light degrees of freedom with $ss$ quantum numbers and unit spin.  The
latter may be regarded as a vector particle, hereafter referred to as a
diquark.  A model for baryon transition amplitudes may then be constructed
by joining external baryons to a quark-diquark loop with specified vertices
and propagators.  (Similar models have been defined for mesons in
\cite{simply} and for $\Lambda_Q$ baryons in \cite{comparison}.)  One of the
main advantages of such a model is that recoil effects are incorporated in
a completely relativistic way in the loop graphs.  An
essential physical effect of soft gluons is to damp out the loop momentum
integrals, and this may be accomplished by including damping factors at the
baryon-quark-diquark vertices.  The model vertices are
chosen to be those shown in Fig. 1.  The Lorentz structure is determined by
heavy-quark symmetry
\cite{Geo}.  Standard propagators with constant masses $m_Q$ and $m$ are used
for heavy quark and diquark, respectively; the propagator for the latter is
taken to be purely
$g_{\mu\nu}$ since the diquark is not a gauge field.  Possible
momentum dependence of the diquark mass is neglected.  The form of the
damping factors at the vertices together with the use of a constant-mass
diquark propagator are the chief assumptions of the model.

In the model, there are no bare propagators for the baryons. The complete set
of kinetic and mass terms for $\Omega_Q$ and
$\Omega_Q^{\ast}$ is obtained from the graph shown in Fig. 2 for the proper
two-point functions (negative inverse propagators) $i\Sigma$ and
$-ig_{\mu\nu}\Sigma_{\ast}$, respectively.  The physical baryon masses
$M_{(\ast)}$ are determined by requiring that
$\Sigma_{(\ast)}(p\!\!\!/=M_{(\ast)})=0$ and
$\Sigma_{(\ast)}^{\prime}(p\!\!\!/=M_{(\ast)})=1$, so that the propagators
have a simple pole and unit residue at the physical masses. The physical
masses are taken to be real. Analysis of the mass functions shows that their
real parts have a single zero which satisfies $m_Q+m < M_{(\ast)} <
m_Q+\widehat{\Lambda}_{(\ast)}$. Note that $m<\widehat{\Lambda}_{(\ast)}$ is
required for the existence of a zero.

The imaginary parts of the mass functions are
nonvanishing for $p^2 > (m_Q+m)^2$.  This means that the baryons of the model
are actually above threshold for decay into a free heavy quark and a diquark.
This is not unexpected, since the particles in the loops propagate freely.
Nowhere have the effects of confinement been introduced.   We will see,
however, that
$\widehat{\Lambda}$ is of order $1500-1600$ MeV; i.e. the characteristic
value of loop momenta which contribute to the mass functions of interest is
large compared with the scale
$\Lambda_{QCD}$ of confinement in QCD.  Therefore, inclusion of the effects
of confinement ought to have only a small effect on the real parts of the
mass functions, while slightly modifying their imaginary parts by shifting
the threshold to a point above $M_{(\ast)}$.  Then it ought to be a
reasonable approximation to simply drop the imaginary parts of the mass
functions entirely.  The same holds for the semileptonic decay form factors;
they all have imaginary parts for the same reason as the mass functions do,
and only the real parts are retained in the model.  It was demonstrated in
\cite{simply} and \cite{comparison} that this procedure is consistent with the
form of the heavy-quark expansion dictated by QCD in the case of mesons and
$\Lambda_Q$ baryons, respectively, that it preserves the Ward identities, and
that it leads to sensible physical predictions.

Heavy-quark symmetries relate properties of $\Omega$ baryons to those of
$\Omega^{\ast}$.  For example, they are degenerate in the heavy-quark
limit.  This can only be true if their damping
factors become the same in this limit, i.e.
$\widehat{\Lambda}$ and $\widehat{\Lambda}_{\ast}$ approach a
common value, say $\Lambda$, and $Z$ and $Z^{\ast}$ approach a common
value, say $A\Lambda$.  The scale $\Lambda$ characterizes the light degrees
of freedom, and can be expected to satisfy $\Lambda\ll m_Q$.
Everything may then be expanded in power series in
$\Lambda/m_Q$:
\begin{eqnarray}
\widehat{\Lambda}_{(\ast)}&=&\Lambda\left\{1+\lambda_{(\ast)}
\frac{\Lambda}{m_Q} +\ldots\right\}  \\
M_{(\ast)}^2 &=&m_Q^2+
c\Lambda m_Q\left\{1+d_{(\ast)}\frac{\Lambda}{m_Q}+\ldots\right\}
\label{mass} \\ Z_{(\ast)}&=&A\Lambda\left\{1+B_{(\ast)}
\frac{\Lambda}{m_Q}+\ldots\right\} \label{norm}
\end{eqnarray}  Equation (\ref{mass}) is the most general relation
consistent with a lowest-order mass difference between baryon and heavy
quark which is common to $\Omega$ and $\Omega^{\ast}$ and is of order
$\Lambda$: $M_{(\ast)}-m_Q\rightarrow \overline{\Lambda} \equiv c\Lambda/2$.

In a graph such as Fig. 2 with one diquark propagator, it is possible to
choose the loop momentum to be the same as the diquark momentum  $k$.  The
damping factors act to suppress contributions of $k$ much larger than
$\Lambda$.  To facilitate the expansion of the heavy-quark propagator, it is
convenient to re-write the light momentum as
$\Lambda k$.  Then the  heavy-quark propagator carries momentum
$p=M_{(\ast)}v+\Lambda k$ with $v^2=1$ and has the expansion
\begin{equation}\frac{-i(p\!\!\!/+m_Q)}{-p^2+m_Q^2}=
\frac{-i}{\Lambda}\left\{
\frac{v\!\!\!/+1}{D} +\frac{\Lambda}{m_Q}\left[F_{(\ast)}
\frac{v\!\!\!/+1}{D^2} + \frac{k\!\!\!/-c/2}{D} \right] + \ldots \right\}
\label{hqexp}
\end{equation} where $F_{(\ast)}=k^2 - c^2/2 + c\,d_{(\ast)}$ and $D = -2 k
\cdot v -c$. The expansion of a vertex factor is
\begin{equation}
\frac{Z_{(\ast)}^2}{-\Lambda^2 k^2+\widehat{\Lambda}_{(\ast)}^2}=A^2
\left\{\frac{1}{D_k}+\frac{\Lambda}{m_Q}\left[\frac{2B_{(\ast)}}{D_k}-
\frac{2\lambda_{(\ast)}}{D_k^2} \right] + \ldots \right\}, \label{vtexp}
\end{equation} where $D_k = -k^2 +1$.  The propagator for light degrees of
freedom is \begin{equation} \frac{ig_{\mu\nu}}{-\Lambda^2 k^2
+m^2} = \frac{ig_{\mu\nu}}{\Lambda^2 D_{\alpha}}  \label{diprop}
\end{equation} where  $D_{\alpha} = -k^2 + \alpha^2$ with
$\alpha=m/\Lambda$.

The constants $c$, $A$, $d_{(\ast)}$ and $B_{(\ast)}$ appearing in
(\ref{mass}) and (\ref{norm}) are fixed by consideration of
the zero and the slope of the mass function as discussed above. To
determine $c$ and $A$, it suffices to compute $\Sigma_{(\ast)}$ at
zeroth order with $v\!\!\!/=1$ using the expansions in (\ref{hqexp}) and
(\ref{vtexp}).  The result is
\begin{equation} \Sigma=\Sigma_{\ast}=\frac{\Lambda}{2} \frac{1}{D_k^2
D_{\alpha} D}, \end{equation}  where we adopt the convention that  an
overall  $4 i A^4 \int d^4k/(2\pi)^4$ is understood in any product of
$1/D$'s, and the real part is implied. The constant $c$ enters this
expression through $D$, and is determined in terms of $\alpha$ by the
on-shell condition
\begin{equation} 0=\frac{1}{D_k^2 D_{\alpha} D}.
\end{equation}  (Note that $\alpha<1$ is required for the existence of a
zero.)  The normalization is fixed by
$\Sigma_{(\ast)}^{\prime} =1$, where the prime denotes differentiation with
respect to $p\!\!\!/$ and is equivalent at zeroth order with $(2/\Lambda)
\partial/\partial c$. The zeroth-order constant $A$ is thus determined in
terms of $\alpha$ by
\begin{equation} 1=\frac{1}{D_k^2 D_{\alpha} D^2}, \label{A}
\end{equation} in which $A$ enters according to the
convention mentioned above.

A similar analysis at first order yields \begin{eqnarray} c \, d_{(\ast)}
&=& \frac{c^2}{2}-1
	+ \frac{1}{D_{k} D_{\alpha} D^2} + \frac{1}{4D_{k}^2
	D_{\alpha}} + \frac{4 \lambda_{(\ast)}}{D_{k}^3 D_{\alpha} D} \label{d} \\
B_{(\ast)}&=& \frac{c}{4} - \frac{F_{(\ast)}}{2D_{k}^2 D_{\alpha}
D^3}+\frac{\lambda_{(\ast)}}{D_{k}^3 D_{\alpha} D^2}
\label{B}
\end{eqnarray}
At order $1/m_Q$, the hyperfine interaction between the heavy and
light quarks moves the $\Omega_Q^{\ast}$ mass up one unit and the
$\Omega_Q$ mass down two units from their common value in the heavy-quark
limit. This is accounted for by writing
$\lambda_{\ast}=g+h$ and $\lambda=g-2h$, where the parameter $g$ contributes
to the common mass and $h$ models the heavy-light hyperfine interaction.
Equation (\ref{d}) shows that the $\Omega_Q^{\ast}-\Omega_Q$ mass
difference at order $1/m_Q$ is proportional to $h$ alone.\footnote{In the
case of mesons, there is an ``intrinsic" splitting independent of $h$ which
arises because a $k^2$ term contributes to the vector meson mass function
and not to the pseudoscalar.  This effect is absent here, since $k$ does
not appear in the numerator of the diquark propagator.}

The main quantities of interest in this paper are the form factors for
the semileptonic decay $\Omega_b \rightarrow \Omega_c^{(\ast)} \ell \nu$,
defined by
\begin{eqnarray} \langle \Omega_c |
	\overline{c}\gamma_{\mu}b | \Omega_b \rangle  &=&
\overline{u}^{\prime} [ F_{1}\gamma_{\mu} + F_{2}v_{\mu} +
F_{3}v^{\prime}_{\mu}] u  \\ \langle \Omega_c |
	\overline{c}\gamma_{\mu}\gamma_{5}b | \Omega_b \rangle  &=&
\overline{u}^{\prime} [ G_{1}\gamma_{\mu} + G_{2}v_{\mu} +
G_{3}v^{\prime}_{\mu}]\gamma_{5} u  \\ \langle
\Omega_c^{\ast} |
	\overline{c}\gamma_{\mu}b | \Omega_b \rangle  &=&
\overline{u}^{\prime}_{\nu} [v^{\nu} (N_{1}\gamma_{\mu} + N_{2}v_{\mu} +
N_{3}v^{\prime}_{\mu}) + g^{\nu}_{\mu}N_4]\gamma_{5} u  \\
\langle
\Omega_c^{\ast} |
	\overline{c}\gamma_{\mu}\gamma_{5}b | \Omega_b \rangle  &=&
\overline{u}^{\prime}_{\nu} [v^{\nu} (K_{1}\gamma_{\mu} + K_{2}v_{\mu} +
K_{3}v^{\prime}_{\mu}) + g^{\nu}_{\mu}K_4] u .
\end{eqnarray}  The notation and conventions of \cite{BB} are used in these
definitions.

In what follows, we will take the limit $m_b \rightarrow \infty$ and
expand the form factors to order $1/m_c$.  The general expressions
deduced in \cite{BB} from QCD using the heavy-quark effective theory are
displayed in Tables 1 and 2, in which
$\xi_i^{\prime} = \xi_i+\varepsilon_c(\widetilde{\xi}_i-3\eta_i)$ and
$\xi_i^{\prime\prime}=\xi_i+\varepsilon_c(\widetilde{\xi}_i+ 3\eta_i/2)
$, respectively.  The expansion parameter is $\varepsilon_c \equiv
\overline{\Lambda}/m_c$.  The definitions of
$\widetilde{\xi}_i$,$\eta_i$ and $\kappa_i$ in Tables 1 and 2 differ from
those in Ref. \cite{BB} by having explicit factors of
$\varepsilon_c$ pulled out to make them universal.

The form factors are computed in the model from the graph shown in Fig. 3, and
are expanded using (\ref{hqexp}), (\ref{vtexp}) and (\ref{diprop}). At zeroth
order, consistency with the form shown in  Tables 1 and 2
is almost trivial, a consequence of the equality of the vertex factors in the
heavy-quark limit and of some standard gamma-matrix algebra.  The
two zeroth-order universal functions of the model are given by
\begin{equation} \xi_1 = \frac{1}{D_k^2 D_{\alpha} D D^{\prime}} ,\;\;\xi_2=0
,\end{equation} where $D^{\prime}= -2 k \cdot v^{\prime} -c$.  The Isgur-Wise
function $\xi_1$ is equal to 1 at zero recoil $v=v^{\prime}$ by virtue of
(\ref{A}).  The structure of the denominator has a simple interpretation:
$D_k^2$ comes from the two equal vertex factors, $D_{\alpha}$ is the
diquark propagator, and $D$ and $D^{\prime}$ are the $b$- and $c$-quark
propagators, respectively.  The Isgur-Wise function is completely determined
by the dimensionless ratio $\alpha$ of the diquark mass to the common damping
scale $\Lambda$ in the  heavy-quark limit, and thus depends only on the
properties of the light degrees of freedom.  In the model, $\xi_1$
is the same function of $\alpha$ and $\omega$ as the single Isgur-Wise
function $\zeta$ determining $\Lambda_b \rightarrow \Lambda_c$ transitions
\cite{comparison}.  However, the numerical value of $\alpha$ is generally
different in the two cases.  We remark also that the Isgur-Wise functions
$\xi_i$ and $\zeta$ for baryons are unrelated to their counterpart $\xi$ for
mesons in the present model \cite{comparison}. (For a model in which they
are related, see \cite{viet}.)

In contrast to the situation at zeroth order, it is a non-trivial
test of any model to verify that the first-order corrections have the form
shown in Tables 1 and 2. There are fourteen form
factors, so there are {\it a priori} fourteen possible independent
universal functions at order
$1/m_c$. However, QCD implies that only seven of these are independent,
and constrains these seven to contribute in the particular pattern shown in
Tables 1 and 2.  Of the seven, two are constrained to vanish at zero
recoil: $\widetilde{\xi}_1(1)=\eta_1(1)=0$.  Explicit computation shows that
the model is indeed consistent with the form allowed by QCD.  The seven
universal functions occurring at first order are found in the model to be
\begin{equation}
\widetilde{\xi}_1=\frac{2\chi_{\ast}+\chi}{3}, \;\;
\widetilde{\xi}_2=\frac{-\xi_1}{3(1+\omega)}, \;\;
\eta_1=\frac{2(\chi_{\ast}-\chi)}{9},  \;\;
\eta_2=\frac{2\xi_1}{9(1+\omega)}, \;\;
\eta_3=0 =\kappa_1=\kappa_2 \label{funs}
\end{equation}
where $\chi_{(\ast)}$ is given by
\begin{equation} \frac{c}{2}\chi_{(\ast)} =
\left[2B_{(\ast)}-\frac{c}{2}\right]\xi_1  +
	\frac{F_{(\ast)}}{D_{k}^2 D_{\alpha} D^2 D^{\prime}}-
\frac{2\lambda_{(\ast)}}{D_{k}^3 D_{\alpha}D D^{\prime}}.
\end{equation} and vanishes at zero recoil by virtue of
(\ref{B}).   In this particular model, there are only two new independent
functions at order $1/m_Q$: $\widetilde{\xi}_1$, which is independent of
the parameter
$h$ representing the hyperfine interaction between heavy and light quarks,
and $\eta_1$, which is proportional to $h$.  Both depend on $\alpha$ as
well.  The consistency with QCD is valid for arbitrary values of these
parameters, without extra constraints.  The model predicts the normalizations
$\widetilde{\xi}_2(1)=-1/6$ and $\eta_2(1)=1/9$, independently of the
parameter values.

It is possible to make rough estimates of $m$ and $\Lambda$, and hence
$\alpha$. In \cite{simply} and \cite{comparison}, reasonable values of
$m_q=250$ and
$2m_q=500$ MeV were found for the nonstrange light quark and scalar diquark
masses, respectively. The light quark hyperfine interaction causes
$\Sigma_c$ to be heavier than $\Lambda_c$ by 170 MeV.   Hence, a reasonable
value for the nonstrange vector diquark mass is
$2m_q+M_{\Sigma_c}-M_{\Lambda_c}=670$ MeV.  A reasonable value
for the strange quark mass was found to be $m_s=400$ MeV \cite{kstargamma}.
For the $ss$ vector diquark mass, it is then natural to take
$m=2m_s+(M_{\Sigma_c}-M_{\Lambda_c})(m_q/m_s)^2=870$ MeV.

In order to estimate $\Lambda$, it is necessary to make use of the mass
functions of the model in unexpanded form. These functions are fully
determined by the diquark mass and the heavy quark mass.
Using the above diquark mass together with the preferred value of
$m_c=1500$ MeV found in the case of mesons \cite{comparison}, the full model
applied to $\Omega_c$ with mass 2711 MeV \cite{E687,ARGUS} yields
$\widehat{\Lambda}=1477$ MeV.  It is not yet possible to find
$\widehat{\Lambda}_{\ast}$ because the mass of $\Omega_c^{\ast}$ is unknown,
but it will be somewhat larger than
$\widehat{\Lambda}$. We expect $\Lambda$ to lie in between, perhaps around
1600 MeV. It is thus reasonable to estimate $\alpha=0.5-0.6$. For
comparison, Table 3 lists values of the analogous parameters for
$\Sigma_c$ and $\Lambda_c$.

Unfortunately, it has not been possible to express the universal
functions of the model in terms of elementary functions of $\omega$.  In
what follows, we give the results of numerical computations.  The Isgur-Wise
function $\xi_1$ is shown in Fig. 4 for
$\alpha=0.5$ and 0.6. Its slope at zero recoil is $-1.02$ and $-1.18$,
respectively. These are less than the upper bound of $-1/3$, valid in the
case where
$\xi_2=0$, which was recently derived from a Bjorken sum rule in \cite{Xu}.
In fact, the model's Isgur-Wise function satisfies $\xi_1^{\prime}(1)<-3/4$
for all values of $\alpha=m/\Lambda$ between 0 and 1 \cite{comparison}.  An
example of a model which does not satisfy the bound derived in \cite{Xu} may
be found in
\cite{EIKL}.

The first-order universal functions $\widetilde{\xi}_1 = x+yg$ and
$\eta_1=zh$ are shown in Fig. 5 for the same values of $\alpha$.  We see that
these functions will be numerically rather small as long as $g$ and $h$ are
not too large. In the case of mesons, it was possible to extract values of the
analogs of $g$ and $h$ from fits to the $B$, $B^{\ast}$, $D$ and $D^{\ast}$
masses \cite{hyperfine}.  The results were $g^{\rm meson}=0.13$ and
$h^{\rm meson}=0.19$.\footnote{The definition of $g$ in \cite{hyperfine}
differs from the present one by a minus sign.}  It is reasonable to suppose
that $g$ and $h$ for baryons are of the same order, so that the first-order
universal functions $\widetilde{\xi}_1$ and $\eta_1$ which are constrained to
vanish at zero recoil remain small across the entire spectrum.

The values of the constants $c$, $A$, $d_{(\ast)}$ and $B_{(\ast)}$, defined
in (\ref{mass}) and (\ref{norm}), are listed in Table 4.  The lowest-order
mass difference is
$\overline{\Lambda}\sim 1300$ MeV.  This compares with $\sim 1000$ MeV in the
$\Sigma_c^{(\ast)}$ system, 790 MeV for $\Lambda_c$ \cite{comparison} and 500
MeV for $D^{(\ast)}$ mesons
\cite{hyperfine}.  We note that the heavy-quark expansion is formally a power
series in $\overline{\Lambda}/2m_c$ \cite{Neurevue}, and its convergence
properties are best in the case of the lowest-lying mesons.  Each set of
hadronic states which are degenerate in the heavy-quark limit has a different
value of $\overline{\Lambda}$, and this increases with the mass of the
hadronic state for fixed heavy-quark mass.  This is true in any model,
because $\overline{\Lambda}\sim M-m_{Q}$.  The $\Omega_c^{(\ast)}$ baryons
have $\overline{\Lambda}/2m_c
\sim 0.4$, and thus may represent the most massive charmed hadrons for which
the heavy-quark expansion makes any sense.

The present model is a specific example of the general physical picture
discussed in \cite{HKKT}, in which current-induced transitions among
heavy-light baryons are described by Bethe-Salpeter bound state wave
functions.  The most general such model treats the light degrees of freedom
as two separate spin-1/2 particles, and represents the amplitudes by
two-loop graphs.  In the heavy-quark limit, two dynamical approximations to
the most general ansatz were discussed in \cite{HKKT}.  In one case, in
which the orbital and spin degrees of freedom of the light quarks are
decoupled, there is a relation between the Isgur-Wise function $\zeta$ for
$\Lambda_b \rightarrow \Lambda_{c}$ and the two for $\Omega_b \rightarrow
\Omega_c^{(\ast)}$.  In the notation of the present paper\footnote{The
functions $f$ and $g$ of \cite{HKKT} are related to the present ones by
$f=2\xi_2$ and $g=\xi_1-(\omega+1)\xi_2$.}, this reads
$\zeta=(2-\omega)\xi_1 +(\omega^2-1)\xi_2$.  In contrast, the present model
 gives $\zeta=\xi_1$ in the artificial case where the
respective values of $\alpha$ are equal, as discussed above.  In the second
case discussed in \cite{HKKT}, the light quarks move independently and the
loop integrals factorize into two pieces.  The two Isgur-Wise functions are
then related by $\xi_1=(\omega+1)\xi_2$, again in contrast to the present
model where $\xi_2=0$ and $\xi_1 \neq 0$.

\newpage
\noindent {\bf Acknowledgement}
\vspace{1ex}

This research was supported in part by the Natural Sciences and Engineering
Research Council of Canada.  The author thanks B. Holdom and Q.P. Xu for many
helpful conversations.

\newpage
\noindent{\bf TABLES}
\begin{equation} \begin{array}{c|ccccccc}
     &    \xi_1^{\prime}   &     \xi_2^{\prime}     &
\varepsilon_c\xi_1 & \varepsilon_c\xi_2  &
\varepsilon_c\eta_3   &   \varepsilon_c\kappa_1   &
\varepsilon_c\kappa_2         \\
\hline F_1  &  \frac{-\omega}{3}  &  \frac{\omega^2-1}{3}  &
    \frac{-\omega}{6}         &       \frac{\omega^2-1}{2}   &
   0     &   \omega+1   &   \omega(\omega+1)   \\  G_1  &  \frac{-\omega}{3}
&  \frac{\omega^2-1}{3}  &
\frac{\omega(1-\omega)}{6(\omega+1)}  &   \frac{(\omega-1)^2}{2}  &
   0     &   \omega-1   &   \omega(\omega-1)   \\   F_2  &  \frac{2}{3}  &
\frac{2(1-\omega)}{3}  &
\frac{-1}{3(\omega+1)}  &        0       &
   1-\omega     &   \omega   &   1   \\ G_2  &  \frac{2}{3}  &
\frac{-2(\omega+1)}{3}  &
\frac{1}{3(\omega+1)}  &        0       &
   \omega +1    &   -\omega   &   -1   \\ F_3  &  \frac{2}{3}  &
\frac{2(1-\omega)}{3}  &
\frac{1}{3}  &      1-\omega       &
   1-\omega    &   -2-\omega   &   -1-2\omega   \\ G_3  &  \frac{-2}{3}  &
\frac{2(\omega+1)}{3}  &
\frac{1-\omega}{3(\omega+1)}  &   \omega-1    &
   -\omega-1     &   2-\omega   &   2\omega-1
\end{array}\label{table1}\end{equation}

\noindent Table 1: $1/m_c$ expansion of
$\Omega_b \rightarrow \Omega_c \ell \nu$ form factors.

\vspace{4ex}
\begin{equation} \begin{array}{c|ccccccc}
     &    \xi_1^{\prime\prime}   &     \xi_2^{\prime\prime}     &
\varepsilon_c\xi_1 & \varepsilon_c\xi_2  &
\varepsilon_c\eta_3   &   \varepsilon_c\kappa_1   &
\varepsilon_c\kappa_2         \\
\hline N_1  &  \frac{-1}{\sqrt{3}}  &  \frac{\omega-1}{\sqrt{3}}  &
    \frac{-1}{2\sqrt{3}}         &       0   &
   0   &   \frac{-\sqrt{3}}{2}   &   \frac{-\sqrt{3}\omega}{2}   \\  K_1 &
\frac{-1}{\sqrt{3}}  &  \frac{\omega+1}{\sqrt{3}}  &
\frac{1-\omega}{2\sqrt{3}(\omega+1)}  &   0  &
   0     &   \frac{-\sqrt{3}}{2}   &   \frac{-\sqrt{3}\omega}{2}   \\   N_2
&  0  &  0  &
\frac{1}{\sqrt{3}(\omega+1)}  &        0       &
   \sqrt{3}     &   -\sqrt{3}   &   \sqrt{3}   \\ K_2  &  0  &  0  &
\frac{1}{\sqrt{3}(\omega+1)}  &        0       &
   -\sqrt{3}    &   \sqrt{3}   &   \sqrt{3}   \\ N_3  &  0  &
\frac{2}{\sqrt{3}}  &  0
  &     0       &
   \sqrt{3}    &   \sqrt{3}   &   -\sqrt{3}   \\ K_3  &  0  &
\frac{-2}{\sqrt{3}}  &   0  &   0   &
   \sqrt{3}    &  \sqrt{3}  &   \sqrt{3}   \\  N_4  &  \frac{-2}{\sqrt{3}}  &
0
 &  \frac{-1}{\sqrt{3}}
  &     0       &
   0    &   0   &   0   \\ K_4  &  \frac{2}{\sqrt{3}}  &  0  &
\frac{\omega-1}{\sqrt{3}(\omega+1)} &   0   &
    0    &   0   &   0
\end{array}\label{table2}\end{equation}

\noindent Table 2: $1/m_c$ expansion of
$\Omega_b \rightarrow \Omega_c^{\ast} \ell \nu$ form factors.

\vspace{4ex}
\begin{equation}
\begin{array}{c|ccccc}
   & M({\rm MeV}) & m({\rm MeV}) & \widehat{\Lambda}({\rm MeV}) &
\Lambda({\rm MeV}) &
\alpha \equiv m/\Lambda \\
 \hline \Omega_c & 2711 & 870 & 1477 & 1600({\rm est.}) & 0.5-0.6 \\
\Sigma_c & 2453 & 670 & 1180 & 1250({\rm est.}) & 0.5-0.6 \\
\Lambda_c & 2285 & 500 & 1010 & 1000 & \sim 0.5 \end{array}\label{table3}
\end{equation}

\noindent Table 3: Comparison of parameter values for various heavy
baryons.

\vspace{4ex}
\begin{equation}
\begin{array}{c|cccc}
 \alpha  & c & A & d_{(\ast)} & B_{(\ast)} \\
 \hline 0.5 & 1.56 & 1.31 & 0.36 + 0.76 \lambda_{(\ast)} & 0.32 + 1.50
\lambda_{(\ast)} \\ 0.6 & 1.64 & 1.16 & 0.40 + 0.70 \lambda_{(\ast)} & 0.32 +
1.80 \lambda_{(\ast)} \end{array}\label{table4}
\end{equation}

\noindent Table 4: Parameters of the heavy-quark expansion for
$\alpha=0.5$ and 0.6.
\newpage
\noindent{\bf FIGURE CAPTIONS} \vspace{6ex}

\noindent {\bf FIG. 1:} Model baryon-quark-diquark vertices.
\vspace{4ex}

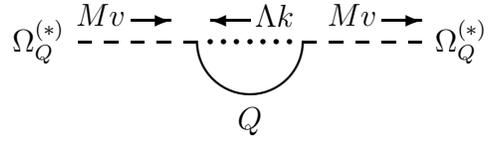
\noindent {\bf FIG. 2:} Graph for the complete set of kinetic and mass
terms (negative inverse propagators) for $\Omega_Q^{(\ast)}$.
\vspace{4ex}

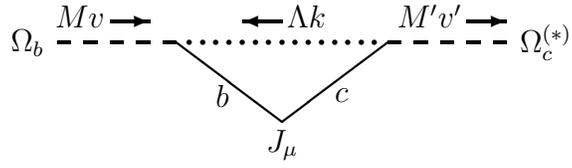
\noindent {\bf FIG. 3:} Graph for $\Omega_b \rightarrow
\Omega_c$ semileptonic decay form factors.
\vspace{4ex}

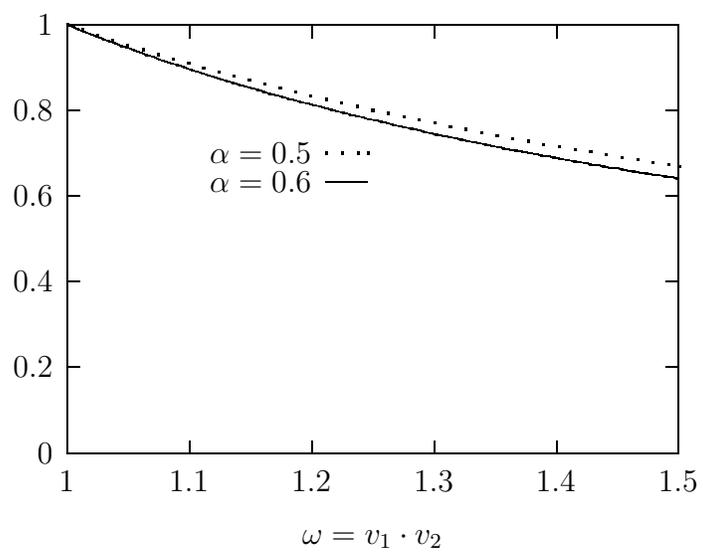
\noindent {\bf FIG. 4:} Model Isgur-Wise function $\xi_1$ for $\alpha=0.5$
and 0.6.
\vspace{4ex}

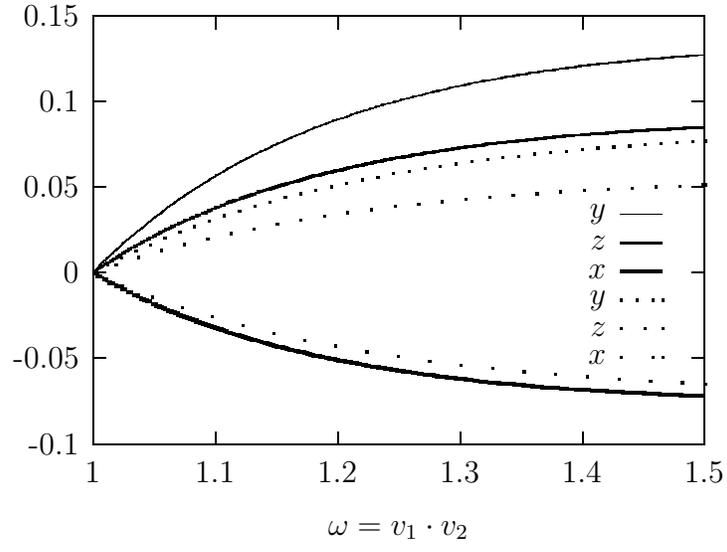
\noindent {\bf FIG. 5:} First-order universal functions $\widetilde{\xi}_1
= x+yg$ and $\eta_1=zh$ for $\alpha=0.5$ (dotted lines) and 0.6
(solid lines).

\end{document}